# Revisiting the Performance of Generative Artificial Intelligence on Introductory Object-Oriented Programming Assessments: Insights from 2026

Marina Lepp*, Joosep Kaimre

Institute of Computer Science, University of Tartu, Estonia

*Corresponding author: marina.lepp@ut.ee, https://orcid.org/0000-0003-3303-5245

**Abstract**

Recent advances in Generative Artificial Intelligence (GenAI) have substantially improved the ability of large language models (LLMs) to generate and explain source code. However, their performance on authentic object-oriented programming (OOP) assessments remains insufficiently understood. This study evaluates five widely used GenAI systems—ChatGPT-5.2, DeepSeek-V3, Gemini 2.5 Flash, Claude Sonnet 4.5, and M365 Copilot—using programming tests and examination tasks from an introductory university OOP course. The generated solutions were assessed using the same grading criteria applied to students and compared with historical student results from the same course, as well as findings from the previous year. Common errors were also analyzed to identify recurring limitations across models. All evaluated GenAI systems achieved higher scores than the average student cohort and frequently obtained full marks on longer programming tasks. Nevertheless, they occasionally produced non-compiling code and continued to struggle with advanced OOP concepts, particularly interfaces, abstract classes, and certain inheritance-related tasks. Performance was also limited on graphics-related questions involving image interpretation. Compared with the previous year, the evaluated systems demonstrated noticeable improvements across most assessments while exhibiting several recurring error patterns. The findings provide an updated evaluation of the capabilities and limitations of contemporary GenAI systems on authentic introductory OOP assessments. They also offer evidence that can inform the design of programming assessments, the responsible integration of GenAI tools into software engineering education, and future studies evaluating the evolution of AI-assisted programming.

## 1. Introduction

Recent advances in Generative Artificial Intelligence (GenAI) have substantially improved the ability of large language models (LLMs) to generate, explain, debug, and transform source code [1–3]. As these systems continue to evolve rapidly, they are increasingly used to support software development and programming-related tasks. Their widespread adoption has also attracted considerable attention in computer science education, where students frequently rely on GenAI systems to assist with programming assignments and coursework [1, 2]. Consequently, understanding the capabilities and limitations of modern GenAI systems on authentic programming tasks has become increasingly important.

A growing body of research has evaluated the performance of GenAI systems in university-level programming courses [1, 4–7], their use as learning support tools [8–11], and their influence on student learning outcomes [12–14]. Although many studies report that GenAI systems can successfully complete introductory programming tasks [5–7], their performance remains inconsistent across different programming topics and assessment types. Previous work has shown that while some models perform at or above the level of typical student submissions, others still struggle with more advanced programming concepts or fail to achieve passing performance on course assessments [15–17]. Furthermore, AI-generated code may contain compilation errors, incorrect implementations, or poor coding practices that can negatively influence novice programmers [18].

Despite the increasing number of studies on GenAI-assisted programming, relatively little attention has been paid to object-oriented programming (OOP). A recent systematic literature review identified only four OOP-focused studies among 125 publications on GenAI in computer science education [19], even though OOP is widely recognized as a threshold concept in computing [20]. Existing work also focuses primarily on English-language prompts, leaving the performance of GenAI systems on programming tasks written in other languages largely unexplored.

Another important challenge is the rapid pace of development of GenAI systems. Conclusions drawn only a year ago may no longer accurately reflect the capabilities of current models. Consequently, longitudinal evaluations using the same assessment tasks are needed to identify how GenAI performance evolves over time and to determine which programming concepts remain challenging despite continuous improvements in model capabilities.

This study provides an updated evaluation of five widely used GenAI systems—ChatGPT-5.2, DeepSeek-V3, Gemini 2.5 Flash, Claude Sonnet 4.5, and M365 Copilot—using authentic assessments from an introductory Java-based object-oriented programming course. Programming tests and examination questions were presented in Estonian without prompt engineering to reflect realistic usage conditions. Historical student results from the same course are used as a reference baseline to contextualize model performance, while comparisons with evaluations conducted one year earlier enable changes in GenAI capabilities to be analyzed over time.

The study is guided by the following research questions:

RQ1. How do GenAI systems perform on authentic introductory object-oriented programming assessments?

RQ2. What recurring error patterns and limitations do GenAI systems encounter when solving introductory object-oriented programming tasks?

This paper makes three main contributions. First, it provides an updated evaluation of five widely used GenAI systems using authentic university programming assessments. Second, it compares the current results with both historical student performance and equivalent evaluations conducted one year earlier, providing a longitudinal perspective on the evolution of GenAI capabilities. Third, it identifies recurring conceptual and implementation errors that continue to challenge current GenAI systems despite their substantial overall performance improvements.

## 2. Background

The rapid development of GenAI has stimulated extensive research evaluating the ability of large language models to solve programming tasks in both software engineering and educational contexts. Recent studies have investigated the accuracy of GenAI systems on programming assignments, examination questions, and software development tasks across a variety of domains, including databases [21], web development [8], and introductory programming [6, 22, 23]. Much of this research has focused on ChatGPT (e.g. [7, 13, 14]), although studies have also evaluated GitHub Copilot [5], Bard [18], and BingAI [24].

To better contextualize these recent developments, it is important to note that the automation of programming tasks has a longer research history. Early work on automated code generation and program synthesis focused on rule-based, template-driven, and specification-guided approaches, typically applied to narrowly defined problems [25]. With the increasing availability of large code repositories, research shifted toward probabilistic and machine learning approaches, leading to the concept of code

“naturalness” and enabling models to learn statistical patterns in source code [26]. Subsequent neural models further improved the ability to capture code structure and usage patterns [27, 28], forming the foundation for modern GenAI systems. The emergence of transformer-based language models has significantly expanded these capabilities by enabling code generation directly from natural-language descriptions. Consequently, evaluating the correctness, robustness, and limitations of these systems on authentic programming tasks has become an important research direction. However, these approaches have largely been evaluated on well-specified benchmark tasks that differ from the open-ended, conceptually demanding problems encountered in educational settings.

In line with this, a number of studies have demonstrated that AI tools frequently perform within the top quartile of student cohorts in introductory programming courses (e.g. [5, 7]). These tools tend to perform well on basic tasks, though their accuracy becomes more inconsistent as the complexity of topics increases [4, 6]. Furthermore, while ChatGPT could successfully complete undergraduate-level coursework, it struggled with postgraduate-level material, performing well on fundamental topics but being outperformed on more advanced ones [7]. In contrast, no significant difference was found between AI performance on introductory versus intermediate tasks, with some evidence that AI tools could succeed on mid-level assessments [23]. GenAI chatbots also performed better on longer programming assignments than on shorter exam-style questions, often failing to provide all correct solutions or misinterpreting implicit requirements [4]. Moreover, Shoufan [16] reported that although ChatGPT received a passing grade, it still underperformed compared to students. However, in certain contexts, ChatGPT failed to pass the exam entirely [17]. Performance differences are also evident across different chatbots. For instance, Bordt and Luxburg [15] observed that ChatGPT-3.5 was able to pass a data structures course but performed below the class average, whereas ChatGPT-4 achieved results similar to those of students. Likewise, it was found that ChatGPT-3.5 scored below average, while Copilot performed on par with students [4]. Furthermore, BingAI performed better than ChatGPT-3.5 in all testing on object-oriented application development [24]. Overall, several studies report that GenAI systems achieve high scores on introductory programming assessments, although their performance becomes less consistent as task complexity increases. The boundary between introductory and intermediate content is often blurred, yet the general trend suggests that while GenAI tools may surpass students in basic coursework, their performance declines on more challenging tasks, though they often remain capable of achieving passing results.

Regarding topic-specific performance, AI tools demonstrated the highest proficiency in algorithms and data structures, followed by operating systems, machine learning, and database management systems [29]. ChatGPT-3.5 also showed the strongest performance in data science tasks, with slightly lower proficiency in operating systems and data network management [24]. Outside traditional university settings, AI chatbots have also been evaluated using competitive programming platforms like LeetCode [30]. ChatGPT outperformed the average human acceptance rate across most problem categories and difficulty levels (easy, medium, and hard) except in tasks requiring bit manipulation. Nonetheless, such challenges differ from university coursework, as participants may be less motivated to perform at their peak given that outcomes do not affect academic grades.

ChatGPT’s performance with non-English input, specifically Czech, has been assessed in information security courses [31]. It successfully passed all four evaluated courses, outperforming the student average in one, while students outperformed it in the remaining three. ChatGPT tended to excel in full-text exams requiring written responses or solutions, but students generally performed better in tasks involving project work, essay writing, and coding small snippets. Furthermore, other research revealed that GPT-3.5’s capabilities in problem generation vary across languages, highlighting the need for further investigation into its performance in a broader range of languages [32].

From an educational perspective, it is also important to consider the challenges faced by human learners. A substantial body of research has examined the difficulties faced by novice programmers, showing that beginners often struggle not only with syntax but also with developing accurate mental models of program behavior and understanding abstract concepts [33, 34]. These difficulties are particularly relevant in the context of object-oriented programming, where learners must reason about abstraction, inheritance, and polymorphism, and often develop only partial or surface-level understanding of these concepts [20, 35, 36]. These well-documented difficulties provide an important point of comparison when evaluating how GenAI systems perform on similar programming tasks.

Research specifically comparing AI assistant performance to student outcomes in Java-based OOP courses remains limited. Studies indicate that while AI tools like ChatGPT can manage simpler coding tasks, they often struggle with more complex problems [37-39]. Nonetheless, they often produce partial solutions that serve as useful starting points for students. Additionally, GenAI chatbots face significant challenges with stack-related tasks and advanced object-oriented concepts such as interfaces, abstract classes, and class hierarchies [4]. Similarly, ChatGPT-3.5 showed major difficulties implementing object-oriented interfaces, generating code that compiled but either failed to function correctly or diverged substantially from the task requirements [24]. Further studies confirm these challenges: AI-generated OOP code frequently has compilation errors, needs multiple prompts to complete all necessary classes and functions, and fails unit tests [18]. Moreover, code produced by ChatGPT is prone to quality problems, including runtime errors, incorrect outputs, and maintainability issues [40]. Another notable limitation is the difficulty chatbots encounter when tasks involve data in the form of UML diagrams or API documentation, as they struggle to fully parse these non-text inputs [38]. This issue is not exclusive to Java or OOP but reflects a broader challenge AI faces in handling non-textual information. Supporting this, Cámara et al. [41] found that ChatGPT struggles to reliably generate UML diagrams and frequently produces syntax errors. Conversely, Hou et al. [22] reported that GPT-4V demonstrated notable skill in solving Parsons problems that include diverse visual representations. Given the limited research directly comparing AI assistants with students, it remains unclear whether AI encounters similar difficulties to students or if it surpasses students in some areas while falling behind in others.

Unlike standardized programming benchmarks, authentic university assessments require GenAI systems to demonstrate both conceptual understanding and practical implementation while addressing learning objectives defined by the course curriculum. Overall, previous studies demonstrate that GenAI systems can successfully solve many university-level programming tasks, although their performance varies considerably across programming concepts and assessment types. Evidence regarding Java-based object-oriented programming remains limited, particularly for the latest generation of GenAI systems and for assessments conducted in languages other than English. Furthermore, relatively few studies have examined how GenAI capabilities evolve over time using comparable assessment tasks. These gaps motivate the present study.

# 3. Methodology

## 3.1 Research context

Object-oriented programming is a widely adopted paradigm centered around objects—instances of classes that encapsulate data and associated behaviors [42]. At the University of Tartu, OOP is taught to first-year students in the “Object-Oriented Programming” course using the Java programming language. Data, tasks, and results from this course form the basis for comparing AI chatbot performance with the performance of novice programmers. While primarily mandatory for Computer Science majors, the course is also taken as an elective by students from other disciplines, making it one of the university’s largest, with 270–330 annual enrollees. Students have to complete an introductory Python course [43] as a prerequisite. Most of the students begin the course without prior experience in Java or object-oriented programming.

The 16-week course follows a flipped-classroom model [44], with weekly lecture videos, quizzes, homework, and practicals. Students complete two programming tests, two group projects [45], and a final exam. The final grade is heavily influenced by the exam (33 points) and tests (2 × 16 points), which together account for nearly 64% of the total score. Practical points are awarded for attendance, and homework assignments include automated feedback, troubleshooters [46] and unlimited resubmissions. Group projects allow creative freedom within set requirements, leading to varied implementations and less consistent grading. Consequently, this study focuses on comparing AI and student performance on the programming tests and exam, which are more standardized.

The tasks used in this study were selected to be representative of the course's core learning objectives and to reflect the types of problems students are required to solve in assessments. Programming test tasks were chosen because they require the implementation of complete programs with multiple interacting components, while exam questions were included to cover a broader range of conceptual understanding. As the tasks were taken directly from actual course assessments without modification, they provide an authentic basis for comparing AI and student performance.

The tests are conducted in weeks 7 and 13, assessing content from the previous weeks. Each test provides a program outline specifying required classes, methods, inheritance, and overall program structure. Students have 105 minutes and may use course materials, but no AI tools or external assistance. While automated tests check for the presence of required components, they do not evaluate the internal logic. Students must test and debug their programs independently to meet the task requirements. The program that does not compile gives zero points. An example of the beginning of the test task (translated into English) is shown in Figure 1.

The test consists of writing a program for renting equipment. The program must meet the requirements below (even if they seem strange). The program must contain the classes Person, RentableEquipment, Laptop, Tablet, and a main class. The main class reads the equipment's data and simulates renting computers. The main class also tests the work of various instance methods. All instance variables of all classes must be private.

1. (3 p) The abstract class RentableEquipment must have private instance variables for the serial number of the equipment (String), the rental status of the equipment (boolean; true, if the computer is in storage), and the borrower's personal identification code (String).
    1. The class must have a three-parameter constructor for setting the serial number, rental status, and borrower's personal identification code. The class must ensure that after setting the serial number in the constructor, it cannot be changed later.
    2. The class must have an abstract void-type method borrow, which takes a Person-type instance as a parameter.
    3. If necessary, get and set methods can be created, for instance variables.
    4. The class must also have a toString method for presenting the equipment information as text, adding a message to the serial number indicating whether the equipment is in storage or rented, and if it is rented, also indicating the borrower's personal identification code.
    5. The RentableEquipment class must implement the Comparable<RentableEquipment> interface, with the compareTo method implemented so that the comparison is performed by serial number.

**Figure 1.** The beginning of programming test 1.

The exam is administered at the end of the course as a computer-based test on Moodle. Students may use course materials, code examples, documentation, and Google search, but they are not allowed to communicate with others, use AI assistants, open IDEs, compile, or run code. The exam lasts 60 minutes and covers the full range of course topics. It begins with an integrity statement worth 1 point and ends with a longer, open-ended task worth 6 points. In between are 13 randomized questions worth 2 points each, including multiple-choice, single-choice, fill-in-the-blank, and matching formats. Once students proceed to the next question, they cannot return to the previous ones. Examples of these shorter questions (translated into English) are shown in the left part of Figure 2. The final task requires explanation and justification. One version asks students to complete a code snippet and explain all valid answers; the other presents faulty code and asks them to evaluate statements about the errors and justify their reasoning. Examples of both types (translated into English) are presented in the right part of Figure 2.

There is a program

```
public ___________ Computer {
    boolean isLaptop();
    int getMemory();
}
```

Which of the following options would fit the gap to make the program correct?

Select one or more:

- ☐ abstract class
- ☐ interface
- ☐ class

There is a program

```
public class Product {
    private double price;
    [      ] Product(double price) {
        this.price = price;
    }
    [      ] double getPrice() {
        return price;
    }
}
```

Fill in the gaps. Each gap can contain one word. The same word goes in all gaps. It is important that the given class and all its methods are accessible from everywhere, including from other packages.

What will be displayed on the screen?

```
Deque<Integer> ad = new ArrayDeque<>();
ad.push(2);
ad.push(3);
ad.push(7);
ad.push(10);
ad.poll();
ad.peek();
System.out.println(ad);
```

Answer: [          ]

Three files are given. Fill in the gaps so that the rest of the code does not need to be changed. A gap can contain a **single word** or **be left blank**. If there are dependencies between any gaps, note them as well (in the style of "if you put ... in the first gap, then you can put ... in the third gap"). If there are multiple options, write down **all the options**. **Explain** your choices!

```
public [        ] [        ] Measurable { //1-2 gap
    public int height();
}
public abstract class Building {
    public [        ] String buildingOwner(); //3 gap
    public [        ] String toString() { //4 gap
        return "Building - owner: " + buildingOwner();
    }
}
public class ApartmentBuilding [        ] Building [        ]
Measurable { //5, 6 gap

}
```

Gap 1-2, explanation:

[          ]

Gap 3, explanation:

[          ]

Gap 4, explanation:

[          ]

The program was given 1 and 2 as command line arguments and expected to get 0.5 on the screen. Unfortunately, this **did not work**. The following list contains various possible reasons. Explain for **each** option how relevant it is in this case. It can be assumed that everything that is necessary has been imported.

```
public class Arithmetic {
    public static void main(String[] args) throws FileNotFoundException {
        OutputStream output = new FileOutputStream("systemout.txt");
        PrintStream printOut = new PrintStream(output);
        System.setOut(printOut);
        System.out.println(f1(Integer.parseInt(args[0]),
Integer.parseInt(args[1])));
    }
    private static double f1(int a, int b) throws ArithmeticException {
        return a / b;
    }
}
```

1. A public method cannot call a private method.

[          ]

2. With the given arguments, the function f1 does not return the number 0.5.

[          ]

3. There is no throws ArithmeticException in the header of the main method.

[          ]

4. The output is written to a file.

[          ]

5. An ArrayIndexOutOfBoundsException is thrown.

**Figure 2.** Examples of the exam tasks: short on the left and long on the right.

### 3.2 Procedure and data analysis

GenAI chatbots were selected based on student-reported usage, gathered through a survey conducted in week 10 via the course's Moodle page. One section of the survey focused on AI assistants, starting with a single-choice question asking whether students had used any AI tools for the course. 87.8% of respondents had used AI assistants in this course at least once. Those who had used AI were then asked which specific chatbots they had used (Figure 3). The five most commonly mentioned systems were selected for further testing. Free versions, including ChatGPT-5.2, DeepSeek-V3, Gemini 2.5 Flash, and Claude Sonnet 4.5, were chosen, as these were assumed to be the most accessible options. The paid version of M365 Copilot was also included, as University of Tartu students receive free access through their institutional accounts, making it a likely choice despite being a paid option. All evaluated GenAI systems were used with their default settings.

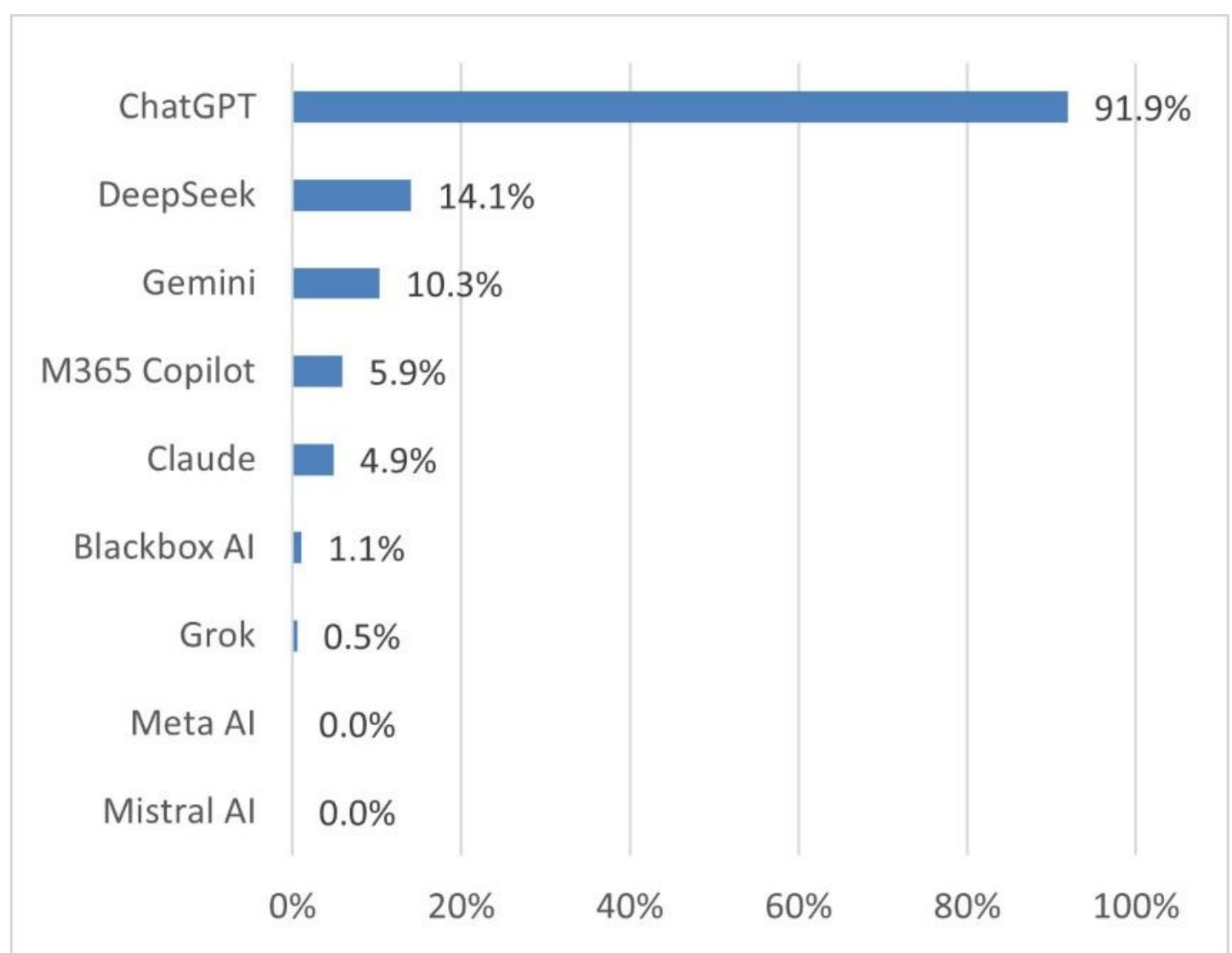

**Figure 3.** Students' reported usage of GenAI chatbots.

To assess how effectively AI chatbots could complete tasks from the introductory OOP course, they were given the full text of the original tests and assignments, with no modifications or translations, meaning all tasks were in Estonian. No additional prompts were used, as the tasks already included clear instructions or questions (e.g. "Write a program..." in programming tests or "Fill in the gap" and "What will be displayed on the screen?" in the exam). No prompt engineering, iterative refinement, or follow-up interactions were performed, ensuring that all evaluated systems received identical inputs. This approach was chosen to reflect a realistic usage scenario in which tasks are often copied directly into chat interfaces, and to minimize potential bias introduced by prompt engineering. However, it is acknowledged that AI performance can be sensitive to prompt phrasing, and alternative formulations or iterative prompting could yield different results. Therefore, the findings should be interpreted as reflecting baseline performance under consistent and controlled conditions rather than optimal AI capability.

Only the first response generated by each GenAI system was evaluated by the course instructor using the course's standard grading scheme. This decision reflects a realistic usage scenario in which users frequently submit assessment tasks directly to GenAI systems and initially rely on the first response. Repeated prompting or iterative refinement was intentionally excluded because it would introduce user-dependent variability and reduce comparability across systems. The evaluation of AI-generated solutions followed the same criteria used for student submissions to ensure comparability. For programming tasks, points were awarded based on the presence of required classes, methods, and functionality, as defined by the course grading scheme. Solutions that did not compile received zero points, consistent with the treatment of student submissions. For exam questions, answers were graded according to predefined correct responses, with partial credit awarded where applicable. Common errors were recorded to identify recurring issues and problematic areas.

Historical student results from the same assessments were used as a reference baseline to contextualize the performance of the evaluated GenAI systems. The purpose of this comparison was not to establish superiority or inferiority between human learners and AI systems, but to provide an interpretable point of reference for authentic university assessments. Since these assessments were originally designed to measure the expected learning outcomes of novice programmers, student performance offers a meaningful context for interpreting the capabilities and limitations of current GenAI systems.

Each assessment task was evaluated once by each GenAI system. Consequently, assessment tasks rather than repeated model outputs constituted the unit of analysis. Each chatbot was tested on three different versions of the course programming tests to increase the number of data points and reduce the impact of task-specific difficulty on results. For the exam, which draws from a Moodle question bank, ten questions from each topic set were selected to measure the AI assistants' average performance across topics. All chatbots received the same set of questions to ensure consistency in comparison. The same programming tests and exam questions were also used in the 2025 evaluation of ChatGPT and Copilot [4], enabling direct longitudinal comparison.

# 4. Findings

## 4.1 Programming test 1

Programming test 1 assessed core topics including Java classes and objects, strings, file handling, lists, polymorphism, interfaces, and abstract, super-, and subclasses. The test was worth 16 points, with 12 points required to pass. The solutions generated by the evaluated GenAI systems were assessed using the course grading scheme. The results are presented in Table 1 together with the corresponding evaluation results for ChatGPT and Copilot from the previous year.

On average, all evaluated GenAI systems achieved scores above the passing threshold, with the exception of Copilot, whose average was affected by one non-compiling solution that received zero points. Among the 285 students who completed the assessment, the average score was 14.61 points. All evaluated GenAI systems, except Copilot, obtained higher average scores than the historical student cohort. ChatGPT also showed improved performance compared with the previous year's evaluation, whereas Copilot's average score decreased because of the non-compiling submission.

**Table 1.** GenAI chatbots' and students' results in the programming test 1.

| | **ChatGPT** | **ChatGPT in 2025** | **DeepSeek** | **Gemini** | **M365 Copilot** | **Copilot in 2025** | **Claude** | **Student average (SD) / median** |
|---|---|---|---|---|---|---|---|---|
| **T1.1** | 15.75 | 14.75 | 15.75 | 16 | 16 | 15.8 | 16 | 13.76 (4.33) / 15.75 |
| **T1.2** | 15.75 | 13.95 | 15.75 | 16 | 16 | 16 | 16 | 14.55 (2.98) / 15.6 |
| **T1.3** | 15.75 | 15.25 | 15.75 | 16 | 0 | 15.75 | 16 | 14.95 (2.68) / 15.7 |
| **Average** | 15.75 | 14.65 | 15.75 | 16 | 10.67 | 15.85 | 16 | 14.61 (3.11) / 15.7 |

ChatGPT and DeepSeek consistently made the same error: both failed to specify the required file encoding. However, ChatGPT avoided a recurring mistake from the previous year, where it incorrectly defined logic within abstract methods in the superclass without overriding them in the subclasses [4]. The Copilot's solution did not compile due to a syntax error in the `toString` method; specifically, a string concatenation that included three consecutive single quotes (`... + "registrinumber='" + getRegistrinumber() + ''' + ...`). This likely resulted from an attempt to enclose the value in single quotes without properly escaping them using a backslash (`'\''`) or switching to double quotes (`"'"`). If the non-compiling error was corrected, Copilot's result would have been 15 points, as the `toString` method in the subclasses was implemented incorrectly by not using the required `super.toString()` method. In contrast, both Gemini and Claude produced error-free solutions.

Some other noteworthy observations, though not actual errors, include that all GenAI chatbots generated complete sets of getter and setter methods, despite the task specifying that they should be created if necessary. Additionally, some chatbots included extra methods not mentioned in the task, whereas students typically follow the instructions more strictly and do not add unasked methods. Since streams were not taught in the course before test 1, students were expected to use the `Scanner` class to read from files. Interestingly, only Gemini did so, while the others used `BufferedReader` and `FileReader`. DeepSeek did not use the `public` access modifier for classes, except for the main class. Notably, both Gemini and Claude included object creation in the `main` method as a fallback in case file reading failed. Furthermore, when implementing the `Comparable` interface, Gemini overrode the `equals` and `hashCode` methods. Among all the AI systems analyzed, Claude stood out for producing particularly nice and detailed program output.

Overall, the evaluated GenAI systems generated highly accurate solutions for the first programming test, with compilation errors occurring only in isolated cases. The most common issues involved omissions of required implementation details rather than incorrect object-oriented program structure.

## 4.2 Programming test 2

Programming test 2 included topics such as streams, exception handling, and data structures, in addition to the material covered in test 1. Like the first test, it was worth 16 points; however, no minimum score was required to pass. The results of grading the GenAI chatbot solutions are shown in Table 2, which also includes comparison data from the previous year for ChatGPT and Copilot. On average, most GenAI tools achieved full or near-full scores, with the exception of Claude, whose average was lower due to one non-compiling solution that received zero points. The average student score for this test was 13.39. All evaluated GenAI systems except Claude obtained higher average scores than the historical student cohort. Compared with the previous year's evaluation, both ChatGPT and Copilot achieved higher average scores.

**Table 2.** GenAI chatbots' and students' results in the programming test 2.

| | **ChatGPT** | **ChatGPT in 2025** | **DeepSeek** | **Gemini** | **M365 Copilot** | **Copilot in 2025** | **Claude** | **Student average (SD) / median** |
|---|---|---|---|---|---|---|---|---|
| **T2.1** | 15 | 11.6 | 16 | 16 | 16 | 15 | 0 | 13.1 (4.06) / 14.8 |
| **T2.2** | 15 | 10.8 | 16 | 16 | 16 | 14.7 | 16 | 13.93 (2.71) / 14.9 |
| **T2.3** | 15 | 14 | 15 | 16 | 16 | 14.9 | 16 | 13.51 (2.89) / 14.5 |
| **Average** | 15 | 12.13 | 15.67 | 16 | 16 | 14.87 | 10.67 | 13.39 (3.59) / 14.75 |

ChatGPT exhibited the same recurring mistake as in the previous year, marking methods as public instead of private. This requirement was outlined in the test, as the methods were not supposed to be accessible outside the class, which likely affected the solution. DeepSeek had the same issue in one of its solutions.

However, ChatGPT correctly handled the tasks involving queues, which had caused recurring errors the previous year [4]. Additionally, the requirement for a non-decreasing sort order did not result in any errors this year. Claude omitted an import statement, resulting in a non-compiling solution. After fixing this minor issue, the solution would have received maximum points. Gemini and Copilot, on the other hand, did not make any mistakes.

Some additional observations, which are not strictly errors, include differences in how sorting was implemented. While students are taught to implement the `Comparable` interface and use the `Collections.sort()` method, all GenAI chatbots used the `Comparator.comparingInt` method along with method references (`::`), which are not covered in the course's core materials. Moreover, the chatbots employed several advanced techniques not typically used by beginner programmers. For example, they computed averages using constructs like: `list.stream().mapToInt(Class::method).average().orElse(0.0);` and created maps using: `list.stream().collect(Collectors.groupingBy(Class::method, Collectors.summingInt(e -> 1)));`. Additionally, methods such as `computeIfAbsent`, `putIfAbsent`, and `removeIf`, which were not presented in the course, were used by several chatbots. There were also some deviations from the task instructions. DeepSeek generated a class implementing the `Serializable` interface, which was not requested. Both DeepSeek and Gemini used an `Integer` type for an instance variable where `int` was expected. Gemini also used a do-while loop for the main program logic, which is an uncommon choice. DeepSeek split parts of the program logic, such as saving to a file, into separate methods. Other minor issues included redundant import statements generated by Gemini, unused methods in the solutions by Gemini and Claude, and additional instance variables created by Copilot that were not specified in the task.

Overall, the evaluated GenAI systems maintained high performance despite the increased complexity of the assessed topics. Most observed issues were related to implementation details or the use of alternative programming constructs rather than fundamental misunderstandings of the assessed concepts.

## 4.3 Final examination

The examination covered the full range of topics addressed in the course. The average scores obtained by the evaluated GenAI systems are presented in Table 3 together with the corresponding results for ChatGPT and Copilot from the previous year's evaluation and the historical student results. Whereas last year ChatGPT's average score fell below the bottom quartile and Copilot's was below the median [4], this year only DeepSeek and Copilot scored within the upper quartile, with the remaining chatbots exceeding it. However, none of the AI tools reached the top 10th percentile.

**Table 3.** GenAI chatbots' and students' average results in the final examination.

| | ChatGPT | ChatGPT in 2025 | DeepSeek | Gemini | M365 Copilot | Copilot in 2025 | Claude | Student average (SD) / median |
|---|---|---|---|---|---|---|---|---|
| **Exam** | 31.01 | 23.82 | 29.5 | 30.73 | 29.49 | 27.13 | 30.95 | 26.9 (3.62) / 27.33 |

The AI assistants all performed nearly flawlessly on questions about objects and classes (Q2 in the Appendix in Table 4). They also answered string-, file-, and list-related questions (Q3 in Appendix Table 4) correctly, except that DeepSeek incorrectly claimed that the expression `"pakend".contains("aken")` is `false`.

Questions about interfaces and abstract classes (Q4–Q5 and Q8–Q9 in Appendix Table 4) proved more challenging. Gemini made fewer errors than the others, but it still incorrectly stated that a class must implement an interface if it merely contains a method with the same name. DeepSeek, Copilot, and Claude made the same mistake. These three also incorrectly assumed that an abstract class is required to implement all methods of an interface. Additionally, ChatGPT, DeepSeek, and Copilot claimed that an abstract class implementing an interface may contain a method with no body and no `abstract` keyword, and that an empty interface is valid even when the implementing class uses the `@Override` annotation. They also showed confusion about when to use `extends` versus `implements`. All AI assistants incorrectly stated that interfaces cannot contain constant variables. Furthermore, ChatGPT and DeepSeek displayed some misunderstandings regarding access modifiers. Finally, every assistant except Gemini claimed that the `abstract` keyword cannot be used within an interface.

While in the previous year ChatGPT and Copilot struggled with class hierarchy [4], particularly with the order of method lookup and how a superclass constructor is called during instance creation, this year all GenAI tools answered the related questions almost perfectly (Q6-Q7 in Appendix Table 4). Only one unusual behavior was observed in Gemini's response, and one unique issue in Copilot's output, though neither was a critical error.

The graphics-related questions included images generated by a JavaFX program and were presented to the GenAI chatbots as a single screenshot. While Claude even offers a *Take a screenshot* feature, Copilot does not accept images alone; some text input is required, so the first line of the question was copied in. DeepSeek accepts image input only when the image contains text. All tools struggled with image recognition, resulting in relatively low scores (Q10 in Appendix Table 4). These were the only questions that received zero points. In cases where the task was to select the appropriate code snippet based on a picture, chatbots sometimes misinterpreted the image. For example, one mistakenly described a black square at the bottom and a red circle at the top, while the actual image showed the reverse. A recurring issue occurred when a code snippet was given, and the chatbot had to choose the correct resulting image. Although the assistants were often able to describe the expected output accurately, they could not generate an image or select the correct one from the screenshot provided. In contrast, GenAI chatbots performed very well on questions about event handling (Q11 in Appendix Table 4). The only exception was ChatGPT, which produced an incorrect answer due to confusion with list indexing, not because of a misunderstanding of the work of events.

Another common error among all GenAI chatbots was the incorrect assumption that data streams cannot interchangeably use `readInt/writeInt` with `readUTF/writeUTF` (Q12 in Appendix Table 4). Regarding the topic of exception handling (Q13 in Appendix Table 4), Claude answered correctly, but ChatGPT and Copilot mistakenly believed that if an exception is thrown within a catch block, it would be automatically caught by the next catch block within the same try statement, which is not the case. Gemini and DeepSeek made unique errors in this topic. While last year the GenAI tools struggled significantly with stack-related questions, giving wrong answers every time [4], this year, all chatbots handled the data structure questions well (Q14 in Appendix Table 4), with the only exception being Gemini, which once returned an incorrect element order for `ArrayDeque`.

The long-form tasks (Q15 in Appendix Table 4) required a comprehensive understanding of the entire course content and revealed a range of mistakes, many of which echoed those from earlier questions. A common issue was that, when asked to provide all valid solutions, AI assistants often gave only one correct answer, for example, suggesting only the `public` keyword despite other valid options, or using just an abstract class or the superclass for object creation. These tasks also required explanations for the chosen solutions, and the assistants generally performed well in this aspect, provided their initial answers were correct. In this question, Copilot struggled more than the other GenAI tools.

Overall, the examination revealed that conceptual object-oriented programming topics—including interfaces, abstract classes, and visual program interpretation—remained more challenging for the evaluated GenAI systems than complete programming tasks. In contrast, topics involving basic programming constructs, data structures, and event handling were handled consistently well.

# 5. Discussion

## 5.1 Performance of GenAI systems

This study evaluated five widely used GenAI systems on authentic introductory object-oriented programming assessments using historical student results as a reference baseline. Across all assessments, the evaluated systems achieved consistently high scores, with most models obtaining higher average scores than the historical student cohort. Overall, the results are consistent with previous studies reporting strong GenAI performance on introductory programming tasks [5–7]. However, they differ from earlier studies, suggesting that GenAI systems perform well on simpler programming tasks but become less reliable as task complexity increases [6, 7]. Instead, the present findings are more consistent with the observations of Savelka et al. [23], who found no substantial performance difference between introductory and intermediate programming tasks. Rather than focusing on direct comparison with students, the results provide evidence of the current capabilities of GenAI systems on authentic university assessments.

One of the most important observations is the rapid evolution of GenAI capabilities. Compared with the previous year's evaluation [4], substantial improvements were observed, particularly across ChatGPT and Copilot. Several recurring errors identified previously—including difficulties with queue-related tasks, inheritance hierarchies, constructor invocation, and object-oriented program structure—were no longer observed. Likewise, ChatGPT improved from achieving examination results below the lower quartile of the historical student distribution in the previous evaluation to exceeding the upper quartile in the present study. These findings also extend the observations of Bordt and Luxburg [15], who reported that ChatGPT-3.5 passed a computer science course but achieved lower scores than the student cohort, whereas ChatGPT-4 performed at a level comparable to students. In the present study, ChatGPT-5.2 obtained higher average scores than the historical student cohort, further illustrating the rapid evolution of GenAI capabilities. These observations demonstrate how quickly conclusions regarding GenAI capabilities may become outdated and highlight the importance of repeated longitudinal evaluations.

Despite these improvements, several limitations persisted across multiple systems. Interfaces and abstract classes remained among the most challenging object-oriented programming concepts, while graphics-related questions continued to produce comparatively low scores because of difficulties in interpreting visual information. These observations are consistent with earlier studies reporting challenges with object-oriented reasoning and UML-related tasks [24, 38, 41], although they contrast with the strong performance reported on visually presented Parsons problems [22]. Interestingly, the evaluated GenAI systems generally performed better on complete programming tasks than on shorter concept-oriented examination questions, suggesting that conceptual reasoning remains more difficult than generating syntactically correct implementations. This observation is also consistent with previous studies [4, 31]. An exception to this trend was ChatGPT, which achieved the highest average score in the final examination while obtaining lower average scores than several other GenAI systems on the programming tests (excluding non-compiling solutions). The reasons for this difference are unclear and warrant further investigation.

Another noteworthy finding concerns multilingual performance. Whereas Malinka et al. [31] reported lower AI performance on Czech-language information security assessments, the evaluated GenAI systems achieved consistently high scores on programming assessments written entirely in Estonian. Although the studies differ in subject area and assessment design, these findings suggest that current GenAI systems are capable of effectively processing programming tasks in languages other than English.

Overall, the results indicate that contemporary GenAI systems have reached a level at which they can reliably solve many authentic introductory object-oriented programming tasks. Nevertheless, recurring difficulties with interfaces, abstract classes, visual reasoning, and concept-oriented questions demonstrate that high overall scores do not necessarily translate into consistent performance on conceptually demanding object-oriented programming tasks. Consequently, while GenAI systems can effectively support routine code generation, solutions involving more complex object-oriented reasoning continue to require careful human verification.

## 5.2 Common errors and limitations of GenAI systems

The second research question focused on identifying recurring errors made by GenAI systems when solving introductory object-oriented programming tasks. Although the evaluated systems achieved high overall scores across the assessments, several common error patterns were observed. These errors were not randomly distributed but clustered around conceptually demanding object-oriented programming topics, interpretation of task requirements, implementation details, and visual reasoning.

The most consistent errors involved interfaces and abstract classes. Several evaluated GenAI systems incorrectly assumed that abstract classes implementing interfaces must provide implementations for all interface methods, confused the use of the `extends` and `implements` keywords, misunderstood the role of the `@Override` annotation, or incorrectly interpreted the use of the `abstract` keyword within interfaces. Difficulties with access modifiers were also observed. Similar challenges have been reported in previous studies [4, 24], suggesting that these concepts continue to represent a persistent weakness of current GenAI systems. While many recurring errors identified in the previous year's evaluation disappeared, misconceptions related to interfaces and abstract classes remained common across several evaluated systems.

A second category of errors involved incomplete interpretation of task requirements. The evaluated systems frequently generated only one valid solution when multiple correct answers were expected, omitted required implementation details such as file encoding specifications, or failed to follow visibility requirements, likely because these were implicitly stated in the assessment. Such behavior indicates that although GenAI systems generally generate syntactically correct and functional code, they do not always identify all explicit or implicit constraints contained in the task description. Similar observations have been reported previously, where AI-generated solutions satisfied the main programming objective while overlooking less prominent requirements [4, 39].

Although comparatively rare, implementation errors were observed as well. These included non-compiling code due to syntax errors or missing import statements, as well as occasional generation of unnecessary classes, methods, or instance variables not requested by the assessment. Previous studies have likewise reported that AI-generated code may contain compilation problems, runtime errors, or maintainability issues despite appearing plausible at first inspection [18, 40]. The relatively small number of such errors in the present study nevertheless indicates considerable progress compared with earlier model generations.

Visual reasoning remained another important limitation. All evaluated GenAI systems achieved noticeably lower scores on graphics-related questions that required interpreting JavaFX output. While current systems are capable of processing images, they frequently misinterpret graphical objects or fail to correctly associate source code with the corresponding visual output. Similar limitations have previously been reported for UML diagrams and other visual software engineering artifacts [38, 41], although they contrast with findings demonstrating strong performance on visually presented Parsons problems [22]. These observations suggest that multimodal reasoning in software engineering contexts continues to require further improvement.

Taken together, these findings indicate that contemporary GenAI systems perform particularly well on tasks involving code generation and recognition of common programming patterns but remain less reliable when solving conceptually demanding object-oriented programming problems or interpreting non-textual information. When the generated solutions were correct, the accompanying explanations were generally well structured and accurate, supporting earlier findings that GenAI systems can produce clear and coherent justifications for their solutions [31]. Interestingly, several recurring error categories closely resemble well-documented misconceptions among novice programmers [33, 34], particularly those related to abstraction, inheritance, and interfaces. This similarity suggests that comparing AI-generated errors with student misconceptions may provide valuable insights into both programming education and the continued development of GenAI systems. Future research should investigate whether these recurring error patterns persist as GenAI systems continue to evolve and whether they can be used to support the design of more effective programming assessments and learning materials.

## 6. Conclusions

This study presented an updated evaluation of five widely used GenAI systems on authentic assessments from an introductory Java-based object-oriented programming course. Across programming tests and the final examination, the evaluated systems consistently achieved high scores, with most obtaining higher average scores than the historical student cohort. Compared with the previous year's evaluation, substantial improvements were observed across nearly all assessed systems. These findings demonstrate the rapid evolution of GenAI capabilities and highlight the value of repeated longitudinal evaluations for tracking changes in model performance over time.

Despite these improvements, several limitations remained consistent across multiple evaluated systems. Interfaces, abstract classes, and graphics-related tasks continued to produce recurring errors, indicating that conceptually demanding object-oriented programming topics remain more challenging than routine code generation. While the evaluated systems generally produced complete and syntactically correct programs, they were less reliable when interpreting implicit task requirements, generating multiple valid solutions, or reasoning about more subtle object-oriented concepts. These recurring error patterns suggest that high overall performance does not necessarily translate into equally reliable performance across all programming topics.

The findings have implications for both software engineering and programming education. From a software engineering perspective, the results demonstrate that contemporary GenAI systems can effectively support routine code generation but still require careful human verification for conceptually demanding object-oriented programming tasks. From an educational perspective, authentic university assessments provide a valuable context for evaluating GenAI capabilities while simultaneously identifying programming concepts that remain difficult for both AI systems and novice programmers.

Analyzing the findings, several limitations should be taken into account. First, AI assistants are highly sensitive to task phrasing and format, and minor changes in wording or input structure can significantly affect performance. A second limitation of this study is its focus on tasks from just one course on Java-based object-oriented programming. Future research should expand beyond Java and OOP to explore whether these patterns persist across different languages and programming paradigms. More work is also needed to assess GenAI performance in non-English educational contexts. Finally, as AI capabilities continue to evolve, longitudinal studies will be essential for understanding how student learning is shaped by ongoing exposure to generative tools.

## Funding

This work was supported by the Estonian Research Council grant "Developing human-centric digital solutions" (TEM-TA120).

## Conflict of interest

The author declares no conflicts of interest.

## Data Availability Statement

The data that support the findings of this study are available upon request from the corresponding author.

# Appendix

**Table 4.** GenAI chatbots' results and mistakes in exam questions.

| | Topic | GenAI | Points | Mistakes |
|---|---|---|---|---|
| Q2 | Objects, Classes | ChatGPT | Avg=2 SD=0 min=2 max=2 | No mistakes |
| | | DeepSeek | Avg=2 SD=0 min=2 max=2 | No mistakes |
| | | Gemini | Avg=2 SD=0 min=2 max=2 | No mistakes |
| | | Copilot | Avg=2 SD=0 min=2 max=2 | No mistakes |
| | | Claude | Avg=2 SD=0 min=2 max=2 | No mistakes |
| Q3 | Strings, Files, Lists | ChatGPT | Avg=2 SD=0 min=2 max=2 | No mistakes |
| | | DeepSeek | Avg=1.89 SD=0.34 min=0.93 max=2 | Did not find the contained substring |
| | | Gemini | Avg=2 SD=0 min=2 max=2 | No mistakes |
| | | Copilot | Avg=2 SD=0 min=2 max=2 | No mistakes |
| | | Claude | Avg=2 SD=0 min=2 max=2 | No mistakes |
| Q4 | Interfaces | ChatGPT | Avg=1.87 SD=0.28 min=1.33 max=2 | Unimplemented method in abstract class without abstract keyword<br>Does not consider that @Override refers to a method |
| | | DeepSeek | Avg=1.55 SD=0.42 min=1.33 max=2 | Unimplemented method in abstract class without abstract keyword<br>Does not consider that @Override refers to a method<br>Class must implement the interface if the method name matches<br>The abstract class must implement the interface methods |
| | | Gemini | Avg=1.93 SD=0.21 min=1.33 max=2 | Class must implement the interface if the method name matches |
| | | Copilot | Avg=1.53 SD=0.47 min=0.67 max=2 | The abstract class must implement the interface methods<br>Does not consider that @Override refers to a method<br>Class must implement the interface if the method name matches<br>Unimplemented method in abstract class without abstract keyword |
| | | Claude | Avg=1.83 SD=0.36 min=1.33 max=2 | The abstract class must implement the interface methods<br>Class must implement the interface if the method name matches |
| Q5 | | ChatGPT | Avg=1.63 SD=0.62 min=0.33 max=2 | Unimplemented method in abstract class without abstract keyword<br>Sorted in the wrong direction with comparable<br>The keyword abstract cannot be used in the interface |
| | | DeepSeek | Avg=1.53 SD=0.63 min=0.33 max=2 | Unimplemented method in abstract class without abstract keyword<br>Sorted in the wrong direction with comparable<br>The keyword abstract cannot be used in the interface |

| | Topic | GenAI | Points | Mistakes |
|---|---|---|---|---|
| | | Gemini | Avg=2 SD=0 min=2 max=2 | No mistakes |
| | | Copilot | Avg=1.5 SD=0.53 min=1 max=2 | Unimplemented method in abstract class without abstract keyword<br>The keyword abstract cannot be used in the interface |
| | | Claude | Avg=1.63 SD=0.48 min=1 max=2 | Sorted in the wrong direction with comparable<br>The keyword abstract cannot be used in the interface |
| Q6 | Class hierarchy | ChatGPT | Avg=2 SD=0 min=2 max=2 | No mistakes |
| | | DeepSeek | Avg=2 SD=0 min=2 max=2 | No mistakes |
| | | Gemini | Avg=1.93 SD=0.21 min=1.33 max=2 | In addition to the correct Error message, it is selected that nothing is printed due to the error |
| | | Copilot | Avg=2 SD=0 min=2 max=2 | No mistakes |
| | | Claude | Avg=2 SD=0 min=2 max=2 | No mistakes |
| Q7 | | ChatGPT | Avg=2 SD=0 min=2 max=2 | No mistakes |
| | | DeepSeek | Avg=2 SD=0 min=2 max=2 | No mistakes |
| | | Gemini | Avg=2 SD=0 min=2 max=2 | No mistakes |
| | | Copilot | Avg=1.93 SD=0.21 min=1.33 max=2 | Changed the end of the answer option and said it was correct |
| | | Claude | Avg=2 SD=0 min=2 max=2 | No mistakes |
| Q8 | Abstract classes | ChatGPT | Avg=1.83 SD=0.36 min=1 max=2 | Used extends with interfaces<br>Subclass cannot widen superclass method access |
| | | DeepSeek | Avg=1.6 SD=0.7 min=0.33 max=2 | Used extends with interfaces<br>Subclass cannot widen superclass method access<br>A private instance variable is accessible in a subclass<br>Subclass cannot redeclare an instance variable |
| | | Gemini | Avg=1.93 SD=0.21 min=1.33 max=2 | Unimplemented method in abstract class without abstract keyword |
| | | Copilot | Avg=1.8 SD=0.32 min=1.33 max=2 | Used extends with interfaces<br>The abstract class must implement the interface methods |
| | | Claude | Avg=2 SD=0 min=2 max=2 | No mistakes |
| Q9 | | ChatGPT | Avg=1.8 SD=0.42 min=1 max=2 | Interfaces cannot contain variables |
| | | DeepSeek | Avg=1.8 SD=0.42 min=1 max=2 | Interfaces cannot contain variables |
| | | Gemini | Avg=1.9 SD=0.32 min=1 max=2 | Interfaces cannot contain variables (int type variable is ok, but String is not) |
| | | Copilot | Avg=1.8 SD=0.42 min=1 max=2 | Interfaces cannot contain variables |
| | | Claude | Avg=1.8 SD=0.42 min=1 max=2 | Interfaces cannot contain variables |
| Q10 | Graphics | ChatGPT | Avg=1.4 SD=0.97 min=0 max=2 | Problems with image recognition. |
| | | DeepSeek | Avg=0.8 SD=1.03 min=0 max=2 | |
| | | Gemini | Avg=0.8 SD=1.03 min=0 max=2 | |
| | | Copilot | Avg=1 SD=1.05 min=0 max=2 | |

| | Topic | GenAI | Points | Mistakes |
|---|---|---|---|---|
| | | Claude | Avg=1.2 SD=1.03<br>min=0 max=2 | |
| Q11 | Events | ChatGPT | Avg=1.95 SD=0.16<br>min=1.5 max=2 | Confusion with list indexes |
| | | DeepSeek | Avg=2 SD=0<br>min=2 max=2 | No mistakes |
| | | Gemini | Avg=2 SD=0<br>min=2 max=2 | No mistakes |
| | | Copilot | Avg=2 SD=0<br>min=2 max=2 | No mistakes |
| | | Claude | Avg=2 SD=0<br>min=2 max=2 | No mistakes |
| Q12 | Streams | ChatGPT | Avg=1.91 SD=0.28<br>min=1.1 max=2 | readInt cannot comprehend input written with the method writeUTF |
| | | DeepSeek | Avg=1.74 SD=0.57<br>min=0.33 max=2 | readInt cannot comprehend input written with the method writeUTF<br>Confusion with the error message |
| | | Gemini | Avg=1.91 SD=0.28<br>min=1.1 max=2 | readInt cannot comprehend input written with the method writeUTF |
| | | Copilot | Avg=1.91 SD=0.28<br>min=1.1 max=2 | readInt cannot comprehend input written with the method writeUTF |
| | | Claude | Avg=1.91 SD=0.28<br>min=1.1 max=2 | readInt cannot comprehend input written with the method writeUTF |
| Q13 | Exception handling | ChatGPT | Avg=1.91 SD=0.28<br>min=1.1 max=2 | Enters multiple catch blocks from a single try |
| | | DeepSeek | Avg=1.9 SD=0.32<br>min=1 max=2 | Adds print statement with exception text |
| | | Gemini | Avg=1.8 SD=0.63<br>min=0 max=2 | a=a causes non-compilation |
| | | Copilot | Avg=1.96 SD=0.13<br>min=1.6 max=2 | Enters multiple catch blocks from a single try |
| | | Claude | Avg=2 SD=0<br>min=2 max=2 | No mistakes |
| Q14 | Data structures | ChatGPT | Avg=2 SD=0<br>min=2 max=2 | No mistakes |
| | | DeepSeek | Avg=2 SD=0<br>min=2 max=2 | No mistakes |
| | | Gemini | Avg=1.8 SD=0.63<br>min=0 max=2 | ArrayDeque elements in the wrong order |
| | | Copilot | Avg=2 SD=0<br>min=2 max=2 | No mistakes |
| | | Claude | Avg=2 SD=0<br>min=2 max=2 | No mistakes |
| Q15 | Question with explana-tions | ChatGPT | Avg=5.7 SD=0.58<br>min=4.33 max=6 | Did not mention all suitable access modifiers<br>Did not use interfaces or abstract classes, only a class<br>Did not mention creating subclass instances with both subclass and superclass types |
| | | DeepSeek | Avg=5.68 SD=0.46<br>min=4.82 max=6 | Did not mention all suitable access modifiers<br>Did not use interfaces or classes, only an abstract class<br>Uses superclass type with a missing method for subclass instance<br>Duplicates the declaration of the variable |
| | | Gemini | Avg=5.73 SD=0.45<br>min=5 max=6 | Did not mention all suitable access modifiers<br>Did not use interfaces, only a class or abstract class<br>Private methods cannot be called from the main in the same class |
| | | Copilot | Avg=5.05 SD=0.89<br>min=3.5 max=6 | Did not mention all suitable access modifiers<br>Did not use interfaces, only a class or abstract class<br>Uses superclass type with a missing method for subclass instance |

| | Topic | GenAI | Points | Mistakes |
|---|---|---|---|---|
| | | | | Duplicates the declaration of the variable<br>Private methods cannot be called from the main in the same class<br>Did not mention creating subclass instances with both subclass and superclass types<br>Unimplemented method in abstract class without abstract keyword |
| | | Claude | Avg=5.58 SD=0.58 min=4.5 max=6 | Did not mention all suitable access modifiers<br>Did not use interfaces or classes, only an abstract class<br>Uses superclass type with a missing method for subclass instance<br>Private methods cannot be called from the main in the same class |